\documentclass[a4paper,twoside]{article}
\usepackage[a4paper,bindingoffset=0.2in,left=2.6cm,right=2.6cm,top=3.3cm,bottom=4.2cm]{geometry}
\usepackage{epsfig}
\usepackage{subcaption}
\usepackage{calc}
\usepackage{amssymb}
\usepackage{amstext}
\usepackage{amsmath}
\usepackage{amsthm}
\usepackage{multicol}
\usepackage{graphicx}
\usepackage{array,multirow,makecell}
\usepackage{hyperref}
\usepackage{pslatex}
\usepackage{apalike}
\usepackage{booktabs}
\usepackage{SCITEPRESS}     

\begin{document}

\title{Toward a multimodal multitask model  for neurodegenerative diseases diagnosis and progression prediction}

\author{\authorname{Lahrichi Sofia, Rhanoui Maryem, Mikram Mounia, El Asri Bouchra}
\affiliation{IMS Team, ADMIR Laboratory, Rabat IT Center, ENSIAS }
\affiliation{Mohammed V University, Rabat, Morocco}
\email{sofia\_lahrichi@um5.ac.ma ,b.elasri@um5s.net.ma ,bouchra.elasri@ensias.um5.ac.ma \\mrhanoui@esi.ac.ma ,mmikram@esi.ac.ma}}


\abstract{
Recent studies on modelling the progression of Alzheimer's disease use a single modality for their predictions while ignoring the time dimension. However, the nature of patient data is heterogeneous and time dependent which requires models that value these factors in order to achieve a reliable diagnosis, as well as making it possible to track and detect changes in the progression of patients' condition at an early stage.
This article overviews various categories of models used for Alzheimer's disease prediction with their respective learning methods, by establishing a comparative study of early prediction and detection Alzheimer's disease progression. Finally, a robust and precise detection model is proposed. \\
}

\onecolumn \maketitle \normalsize \setcounter{footnote}{0} \vfill

\section{\uppercase{Introduction}}
\label{sec:introduction}
Alzheimer's disease (AD) is a progressive, irreversible neurodegenerative disease characterized by an abnormal build-up of amyloid plaques and neurofibrillary tangles in the brain, resulting in memory, thinking and behavior issues. AD is the most common form of dementia characterized by a slow and asymptomatic progress of the disease.\newline
AD is clinically very heterogeneous, varying from patient to patient in terms of cognitive symptoms, test results, and rate of progression. Indeed, several recent therapeutic trials have shown variable efficacy from one subset of patients to another. Currently available treatments only slow the progression of AD, and no definitive cure has been developed yet.
When it comes to AD, patient data \cite{zhang2012multi} is heterogeneous, but complementary, of different types; magnetic resonance imaging (MRI), positron emission tomography (PET), genetics, cerebrospinal fluid (CSF), etc. 
The combination of multimodalities \cite{weiner2013alzheimer} facilitates the detection of changes in the patient's states and constitutes a reliable diagnosis.\newline
Patient data is gathered from different visits and from continuous patient monitoring. The state of the disease at any given time is not independent of the condition at a previous time. Therefore, AD data are not only multimodal but could also be considered as time series and longitudinal series.\newline
In the practical diagnosis of AD \cite{jo2019deep}, the most widely used neuroimaging dataset comes from the Alzheimer's Disease Neuroimaging Initiative (ADNI) which contains \cite{jack2008alzheimer} socio-demographic data (gender, education level), the APOE genotype and five neuropsychological test results: MMSE (mini-mental state examination), CDR-SB (the sum of boxes of clinical dementia rating), ADASCog (Alzheimer's Disease Assessment Scale cognitive sub-scale), LMT (Logical Memory Test ) and RAVLT (Rey Auditory Verbal Learning Test).\newline
Several learning methods have been developed for the classification and prediction of MCI (Mild Cognitive Impairment) to AD conversion, namely Machine Learning and Deep Learning.\newline
After extensive bibliographical and analytical work on various articles, we will give an overview of the categories of approaches used for the prediction of Alzheimer's disease, then we will study the different learning techniques for each approach and establish a comparative table of this last.
This categorization was essential to us because it will allow us at the end of the study to propose a stable and precise detection model.\\
The remainder of the article will consist of three main parts; the first will be the most important because through it, we will establish the bases of our study by defining the concepts of multimodal and multitask learning.
The second part will be devoted to the comparative study of the different models for predicting the progression of AD.
And in the last part, we will propose a model grouping together all the adequate criteria for a better detection of the disease.

\section{\uppercase{Background}}

In this section, we introduce the main notions necessary to understand the context of our proposal.
\subsection{Multi-modal learning}
Multimodal learning combines \cite{zhang2020survey} data from multiple sources that are semantically correlated and sometimes provide complementary information to each other, resulting in more robust predictions.\newline
The multimodal learning model not only captures the correlation structure between different modalities but also allows to recover missing modalities given those observed.\newline
The multimodal learning model for example \cite{intro2020} can combine two deep Boltzmann machines, each corresponding to a modality. An additional hidden layer is placed on top of the two Boltzmann machines to give the common representation. \newline
Multimodal learning is divided into four stages: \newline
(1) Representing inputs and summarize data in a way that expresses multiple modalities.\newline
(2) Translating (map) the data from one modality to another.\newline
(3) Extracting important features from information sources while creating models that best match the type of data.\newline
(4) Fusion and co-learning. Consists in combining the information of two or more modalities to make a prediction. This combination should be normally weighted. \newline
 \subsection{Multitask learning}
Multitask learning (MTL) aims to extract useful information from multiple tasks and use their correlations to help learn a more accurate pattern for each task.\newline
Multitask learning has been used successfully in all applications of machine learning, natural language processing\cite{worsham2020multi}, speech recognition\cite{pironkov2016multi}, computer vision, and drug discovery. \newline
MTL \cite{zhang2018overview}can be categorized into several categories, including supervised multitask learning(MTSL), unsupervised multitask learning, semi-supervised multitask learning... \newline
MTSL aims to learn n functions for the n tasks of the labeled training set, (Each task can be a classification or regression problem) after it uses fi (•) to predict the labels of the data instances i from the j th task.
There are three categories of MTSL patterns to indicate the relationship between stains:\newline
\begin{bf}Feature-based MTSL.\end{bf} Which learns common features of different tasks in order to share knowledge and avoid using original representations directly. It can easily be affected by outliers.\newline
\begin{bf}Parameter-based MTSL.\end{bf} Which finds how the model parameters of different tasks are related, which leads to ranking. It is more robust to outliers.\newline
\begin{bf}Instance-based MTSL.\end{bf} Which aims to find in one task the instances useful for other tasks. It is not used too much.\\
In unsupervised multitask learning, each task, can be a clustering problem, aims to identify useful patterns contained in a training dataset composed only of data instances.\\
In multitask semi-supervised learning, each task aims to predict the labels of data instances from the labeled and also unlabeled data.

\section{\uppercase{Comparative Study}}
\subsection{Models categories}
After an in-depth bibliographical study on the progression of AD, and after consulting a large number of articles that seemed relevant to us, two main ideas emerge: Research on modeling disease progression has received special attention from researchers around the world, and the modeling has been divided into four categories. 
\newline \newline
\hspace*{0.25cm} \begin{bf}Mono-modal mono-task learning.\end{bf} Where the model optimizes only one objective function based on a type of data. Through this model, neither the correlation between the tasks, nor the collective information between the modalities are explored.\newline
\hspace*{0.25cm} \begin{bf}Mono-modal multitask learning.\end{bf} Where the tasks share a few training instances. The relationship between the spots is modeled assuming that they share a common representation space or they share certain parameters. However, they don't take into account the relationship between the different modalities of the same task.\newline
\hspace*{0.25cm} \begin{bf}Multi-modal mono-task learning.\end{bf} Where several modalities are taken into consideration in the prediction of a state while ignoring information from other tasks.\newline
\hspace*{0.25cm} \begin{bf}Multi-modal multitask learning.\end{bf} Where each task of the problem has characteristics from several modalities and or several tasks are linked to each other in a chronological sequence. \newline
Artificial intelligence is advancing in understanding the world around us and the operating model that most closely matches our real environment and provides greater precision is multitask multimodal learning \newline

\subsection{Related works}
In our bibliographical study, we focused on the following articles which seem to us the most relevant and which meet our expectations. \\\\
\hspace*{0.25cm} \begin{bf}Mono-modal mono-task learning\end{bf}\\
\cite{cui2019rnn} proposed a single-task, monomodal model based on six-step MRI time series data for the detection of AD progression. They use a stacked CNN-BGRU (Convolutional  Neural  Network Bidirectional Gated Recurrent Unit) pipeline. It achieves an accuracy of 91.33\% for AD vs CN (normal cognitive), and 71.71\% for pMCI (progressive MCI) vs sMCI (stable MCI). However, relying on MRI alone is insufficient in the medical field.\\\\
\hspace*{0.25cm} \begin{bf}Mono-modal multitask learning\end{bf}\\
\cite{liu2018joint} proposed a deep multi-channel multi-task convolutional neural network for classification (Multi-class classification) and regression using MRI data and BG demographic information.\\
\cite{lopez2018advances} have developed a new nonlinear multitasking (three tasks: animal naming (AN), picture description (PD) and spontaneous speech (SS)) approach based on automatic speech analysis. They introduced linear features, perceptual features, Castiglioni fractal dimension and Multiscale Permutation Entropy into their analysis.\\
Based on the MRI data, they performed a classification, using Multilayer Perceptron (MLP) and Deep Learning by means of Convolutional Neural Networks (CNN) (biologically- inspired variants of MLPs) which led to promising results.\\\\
\hspace*{0.25cm} \begin{bf}Multi-modal mono-task learning\end{bf}\\
\cite{pan2018synthesizing} proposed a two-step deep learning framework for the diagnosis of AD using both MRI and PET data. in the first step, they assign the corresponding MRI data to the missing PET data using 3D-cGAN (3D Cycle-consistent Generative Adversarial Networks) to capture their underlying relationship. In the second step, they develop LM3IL (Landmark-based Multi-modal Multi-Instance Learning Network) which learns and merges the characteristics necessary for the diagnosis of AD and the prediction of MCI. \\
Using the ADNI dataset, \cite{shi2017nonlinear} proposed a method of transforming nonlinear feature space into more linearly separable data using SVM. They chose TPS (Thin-platespline) as a geometric model because of its power of representation. They also adopted a feature fusion strategy based on a deep network by SDAE (Denoising Sparse Auto-Encoder) to merge the transverse and longitudinal features estimated from MRI brain images. \\
\cite{lee2019predicting} proposed a one-task multimodal deep learning approach by incorporating multi-domain longitudinal data. They applied a GRU (Gated Recurrent Unit) for each modality (4Gru) to produce feature vectors of fixed size which will be concatenated to form an input for the final prediction where regularized logistic regression is used for the classification of MCI- C and MCI-NC (MCI converter and MCI non-converter).
The model on a very small number of features (no addition of BG data) has better prediction, it only optimizes binary classification tasks.\\\\
\hspace*{0.25cm} \begin{bf}Multi-modal multitask learning\end{bf}\\
\cite{lahmiri2019performance} proposed M3T multimodal multitask model for the prediction of AD over two years. The latter has two main stages: (1) Selecting multitask features using the Lasso (2) Using an SVM (Support Vector Machine) model for separate classification and regression.\\
Multi-source multitask learning (MSMT) simultaneously considers two types of prior knowledge.
1) Source consistency
2) Slow temporal evolution\\
\cite{nie2016modeling} proposed a linear MSMT model that predicts the future disease status over 2 years (M06-M48) of new patients, based on their health information at the first moment (Baseline).\\
\cite{tabarestani2020distributed} propose a distributed multitask multimodal model to predict MMSE cognitive measures of AD progression, the latter individually exploits several multitask regression coefficient matrices for each modality, then It concatenates the risk factors with the predicted y of each modality then it goes through gradient boosting to group the results of different modalities and reduce their prediction error.\\
\cite{nie2015beyond} proposed an adaptive  multimodal multitask linear learning model (aM2L) to regularize the modality agreement for the same task, the temporal progression on the same modality and the weight of the modalities.\\
\cite{nie2016modeling} and (Nie  et  al.,  2015) in their predictions, don't incorporate the follow-up observations. Example: To predict her condition at M24 (24 months after Baseline), they don't merge the observations of Baseline, M06 AND M12.\\
\cite{li2015robust} presented a robust deep learning system to identify the different stages of progression of patients with AD based on MRI and PET. They used the dropout technique to improve classical deep learning by preventing its weight co-adaptation, which is a typical cause of deep learning overfitting. They stacked multiple RBMs (Restricted Boltzmann machine) to build a robust deep learning framework, which integrates stability selection and multitask learning strategy.\\
\cite{el2020multimodal} proposed a robust ensemble deep learning model based on a stacked convolutional neural network (CNN) and a bi-directional long-term memory network (BiLSTM). This multimodal multitask model jointly predicts several variables based on the fusion of five types of multimodal time series data plus a basic knowledge set (BG). The predicted variables include the multi-class task and four critical cognitive score regression tasks. This model gave equal weights for the classification and regression tasks. \\\\
The table below \ref{Tab:Models} is a summary of some works seen previously, it lists the advantages and limitations of its different models.
\subsection{Synthesis of models}
All the models seen previously for the study of the prediction of the progression of AD, use the ADNI database which can contain missing data. In order to overcome this problem, among the methods that were used:\\
DEL: We simply eliminated the subjects with either missing sources or missing labels.\\
ZERO: We assigned zero value to any element that is missing.\\
 KNN: The k-nearest neighbor (KNN) method replaced the missing value in the data matrix with the corresponding value from the nearest column ...\\
In recent years, machine learning algorithms (such as SVM and random forest, SDAE) and deep learning (CNN, GRU, LSTM, BGRU, BiLSTM,...) have been used for the design of a predictive model of AD progression.\\
And for the generalization of these models to avoid overfitting ,among the regularizers that have been used ,we find: lasso ,sparce group lasso,l2,1 norm, ridge regression ,fused group lasso(convex and non-convex),dropout ......\\

\cite{liu2013multiple} and  \cite{duchesne2009relating} used regular machine learning techniques to study multimodal single-task classification and regression, respectively.\\
To detect AD progression based on a multimodal single-task deep learning model, we find; \cite{spasov2019parameter} who proposed a classification model based on a CNN ,\cite{lee2019predicting} who also proposed a binary classification model based on GRU and used logistic regression for regularization.\\
In a real medical environment, many modalities are analyzed and multiple clinical variables must be predicted. Multimodal single-task models will not provide in this case sufficient information for the study of AD progression, hence the interest of the multimodal multitask model.\\
This model has been used by \cite{lahmiri2019performance} for classification and regression using SVM and SVR respectively. In order to generalize their model, they used the Lasso regularizer.\\
It was also used by el sappagh who added to their model the time series constraint which is consistent with the longitudinal nature of AD because the patient's state at a given time is not independent of his state at an earlier time.\\
They used the CNN-BILTSM for multiclass classification and for regression of 4 cognitive scores. For the regularization of their model, they opted for the dropout. \\
\cite{el2020multimodal} consider that the modalities have the same weight of importance which is not always true in the medical field. \\
\begin{table*}[!]
\small
  \centering
  \caption{Comparative table of models for predicting the progression of Alzheimer’s disease}
\scalebox{0.85}{
\begin{tabular}{ | m{4em} | m{4em}| m{5em} |m{3em} |m{4em} |m{11em} |m{5em} | m{5em} |} 
\toprule
\textbf{Study}                                     & \textbf{Data}                             & \textbf{Modality}               & \textbf{Merging} & \textbf{Time-series} & \textbf{Performance}                                                                                                                                & \textbf{Model}   & \textbf{Task}                                    \\ \midrule
\cite{lee2019predicting}          & ADNI \cite{databaseadni} & Demographic, MRI, CSD, CSF(LCR) & NO               & YES (4 Steps)         & Accuracy: 81\% (MCI/AD)                                                                                                                                    & GRU       & Classification                                 \\ \midrule
\cite{lahmiri2019performance}     & ADNI                                      & MRI, FDG-PET, LCR               & NO               & NO                   & Classification Accuracy: 93,3\%(CN/AD), 83,2\%(CN/MCI), 73,9\% (sMCI/pMCI). Regression Accuracy: 0,697(MMSE), 0,739(ADAS)                            & SVM      &Classification and regression                                   \\ \midrule
\cite{nie2016modeling}            & ADNI                                      & MRI,PET, CSF,PROPT, META        & NO               & NO                   & ADAS-Cog =90,94\% MMSE=87,94\%                                                                                                                      & MSMT                &Regression                       \\ \midrule
\cite{cui2019rnn}                 & ADNI                                      & MRI                             & NO               & YES(6 Steps)         & Classification: Accuracy: 91,33\% (AD/ NC), 71,71\% (pMCI:progressive MCI / sMCI:static MCI)                                                           & Stacked CNN-BGRU    &Classification                        \\ \midrule
\cite{tabarestani2020distributed} & ADNI                                      & MRI, PET, COG, CSF              & YES              & NO                   & MMSE: 70.1\%                                                                                                           & distributed multimodal multitask learning   &Regression                                     \\ \midrule
\cite{liu2018joint}               & ADNI                                      & MRI                             & YES              & NO                   & Classification Accuracy: 51,8\% (CN/sMCI/pMCI/AD). Regression: (CDRSB, ADAS11, ADAS13, MMSE): 1.666, 6.2, 8.537, 2.373                                & CNN           &Classification and regression                             \\ \midrule
\cite{shi2017nonlinear}           & ADNI                                      & MRI, Age                        & YES              & NO                   & Classification: TML(Theoretic Metric Learning)-SVM Accuracy:AD/NC:91.95\% MCI/NC:83,72\% Multi-modal S-DSAE Accuracy:80.91\%(MCI/NC) 88.73\%(AD/NC) & TML-SVM SDAE (Stacked Denoising Sparse AE) &Classification \\ \midrule
\cite{pan2018synthesizing}        & ADNI                                      & MRI, PET                        & YES              & NO                   & Classification Accuracy: 92,5\%(HC/AD) 79,06\%(sMCI/pMCI)                                                                                           & 3D-CNN+GAN LM3IL  &Classification                         \\ \midrule
\cite{li2015robust}               & ADNI                                      & MRI,PET, CSF                    & NO               & NO                   & Classification Accuracy: 91.4\%(AD /HC), 77.4\%(MCI/HC), 70.1\% (AD/MCI),and
57.4\%(MCI.c/MCI.NC)                                                                                                           & RBM  &Classification                                       \\ \midrule
\cite{nie2015beyond}              & ADNI                                      & MRI, PET, CSF, PROPT, META      & NO               & NO                   & MMSE:89,01 \%ADAS-Cog: 91,68\%                                                                                                                      & aM2L &Regression                                      \\ \midrule
\cite{el2020multimodal}           & ADNI                                      & MRI, PET, CSD, ASD, NPD         & YES              & YES(15 Steps)        & ACC: 92.62\%, PRE:
94.02\%, F1: 92.56\%,
REC: 98.42. MAE:
0.107, 0.076, 0.075, and
0.085, (FAQ, ADAS,
CDR, MMSE)                                                                                                                                  & StackedCNN-BiLSTM   & Classification and regression                       \\ \bottomrule
\end{tabular}
}
\label{Tab:Models}
\end{table*}

\section{\uppercase{Proposed Model}}
Alzheimer's disease (AD) is a longitudinal disease, that is to say the state of a patient at a time t depends strongly on his state at t-1. AD cannot be triggered spontaneously; there must be a prior appearance of signs that predict the manifestation and evolution of the disease. Most models studied in the previous section do not take this notion into consideration.\\
Several factors can influence the diagnosis outcome of AD, and even the most indirect symptoms may have in some cases an effect on the detection of AD, hence the interest of introducing Multitask Learning. Multitask Learning will allow us to find a relationship between the features, the instances and the parameters involved in the detection of the disease.\\
Moreover, patients’ data is considered multimodal since they’re gathered from various sources such as MRI, PET, CSD, ASD, NPD etc... These modalities can also interact with each other, because the information extracted from a modality can be complementary to other modalities, which opens the possibility of improving the performance of the model for AD progression prediction. Additionally, it is important to note that the data retrieved via different modalities may vary in terms of importance. For instance, MRI and PET are considered a more reliable and relevant data sources, hence the interest in introducing the idea of a weighted model. \\
According to the results of the comparative study seen in the previous section, which follows closely the same direction mentioned previously, we can conclude that the most adequate, precise and stable model is the Multimodal Multitask Model based on time series. \\
The research axis that can be developed from this study, will center around a Multimodal Multitask system based on time series, which will be able to simultaneously regularize the modality weighting, the temporal progression as well as the modality agreement, i.e. the status of the patient estimated by different modalities must be consistent. The model will also be capable of taking into consideration the relationship between AD progression and the patient's comorbidities (cardiovascular disease, depression, genitourinary renal metabolism, endocrine, etc.). Such model has not yet been addressed in the literature. \\
To advance on this research axis, we propose a Machine Learning or a Deep Learning architecture, multitask, multimodal (using MRI, PET, CSD, ASD, NPD), trained on the ADNI database and the patients' demographic data. This model predicts AD progression status of a through a multiclass classification task, and the values of two cognitive scores ADAS, MMSE which will be implemented as two regression tasks.\\
In order to meet the different expectations as well as possible, two possible scenarios seem to open up at the moment:\\
-	The first approach consists in extracting separately the temporal characteristics of each modality (MRI, PET, CSD, ASD, NPD), then merged with the demographic data to extract the common characteristics that respond to each task by applying a multitask learning. \\
-	Or by applying multitasking to each modality and merging the initial results with demographic data, which is assumed to be a time-invariant information. This model has the ability to stop the propagation of an error from one modality to another.\\
The loss function in both approaches will include a first term for classification, a second term for regression and 2 last terms to regularize the weight and the agreement of the modalities.\\

\section{\uppercase{Conclusion}}
\label{sec:conclusion}
Alzheimer's disease \cite{thung2017multi} is currently calling the attention of many researchers. A considerable amount of effort is being put to understand the biological and physiological mechanisms of AD as well as its monitoring and early detection.
The presented works in this paper have raised several approaches for the detection of AD, namely Monomodal monotask, Monomodal multitask Multimodal monotask and Multimodal multitasking.
Following our comparative study of different works and papers, we believe that a major research axis has been revealed in which the Multimodal Multitask based on time series could at the same time to regularize the modality weighting, the temporal progression, the modality agreement and which will even take into consideration the relationship between the progression of AD and the patient's comorbidities. This model could bring a considerable advance in the field of medical research. \\
\vfill
\newpage
\bibliographystyle{apalike}
{\small
\bibliography{MM}}

\end{document}